\documentstyle[10lomcon,cite,epsfig]{article}

\begin{document}

\title{ANOMALOUS COUPLINGS AT LEP2}

\author{ D.Fayolle \footnote{e-mail: fayolle@clermont.in2p3.fr}}

\address{Laboratoire de Physique Corpusculaire}
\address{Universit\'e Blaise Pascal IN2P3/CNRS}
\address{24 avenue des Landais, 63177 Aubi\`ere Cedex, France \\
$\;$ \\}

\maketitle\abstracts{
In its second phase, LEP has allowed to study four fermion processes never observed before.
Results are presented on the charged triple gauge boson couplings (TGC) from the $W$-pair, Single $W$ and Single $\gamma$ production.
The anomalous quartic gauge couplings (QGC) are constrained using production of $WW\gamma$, $\nu\bar{\nu}\gamma\gamma$ and $Z\gamma\gamma$ final states.
Finally, limits on the neutral anomalous gauge couplings (NGC) using the $Z\gamma$ and $ZZ$ production processes are also reported.
All results are consistent with the Standard Model expectations.}

\section{Introduction}

The Large Electron-Positron (LEP) collider has been running above the $W$-pair production threshold since 1996 until the LEP stopped in 2000, at centre-of-mass energies between 161 GeV and 208 GeV.
This has allowed each of the four experiments ALEPH, DELPHI, L3 and OPAL to collect nearly $700 \; pb^{-1}$ of data.

The non-abelian structure of the Standard Model (SM) leads to three and four charged gauge boson vertices of which the couplings are specified in the bosonic part of the lagrangian.
The self-interactions of these bosons correspond to the vertices $\gamma WW$, $ZWW$ and four quartic $WWXX^\prime$, where $XX^\prime$ is either $WW$, $\gamma \gamma$, $ZZ$ or $Z\gamma$.
Even though the SM is experimentally so reliable, there are still some theoretical problems if one looks at higher energies. One way to cope with this is to consider the SM as an effective theory and assume that New Physics (NP) exists at an higher energy scale, inducing deviations of physical observable values from the SM predictions.
The anomalous gauge boson couplings are thus introduced in the upper-dimensional lagrangian.

The study of the charged TGCs is presented in the next section, and the constraints on QGCs are reported in section~\ref{sec:qgc}. The NGCs are described in section~\ref{sec:ngc}. Conclusions are given in section~\ref{sec:conclusions}.

\section{Charged TGCs}\label{sec:tgc}

\subsection{$W^+W^-$ channel}

One of the most important SM process at LEP2 energies is the $W^+W^-$ production because it allows to measure the $W$ mass and the charged triple gauge boson couplings.

In addition to the $t$-channel $\nu$-exchange, $W$-pair production in $e^+e^-$ annihilation involves the triple gauge boson vertices $WW\gamma$ and $WWZ$ which are present in the SM due to its non-abelian nature.
The most general Lorentz invariant lagrangian which describes the triple gauge boson interaction involving W bosons has fourteen independent terms, seven describing each $WWV$ vertex, with $V = Z , \gamma$.
Assuming electromagnetic gauge invariance, C, P and CP conservation, a total of five couplings remain, which are $\Delta g_1^Z$, $\Delta \kappa_\gamma$, $\lambda_\gamma$, $\Delta \kappa_Z$ and $\lambda_Z$~\cite{lep2}~\footnote{$\Delta$ denotes the difference of these couplings with respect to the SM value.}.
At the tree level in the SM, $g_1^Z = \kappa_\gamma = \kappa_Z = 1$, while $\lambda_\gamma = \lambda_Z = 0$.
Requiring $SU(2)_L \times U(1)_Y$ leads to three independent couplings~\footnote{The common set used includes $g_5^Z$ which is C- and P-violating.}, $\Delta g_1^Z$, $\Delta \kappa_\gamma$ and $\lambda$, related through $\Delta \kappa_Z = \Delta g_1^Z - \Delta \kappa_\gamma \tan^2 \theta_W$ and $\lambda_Z = \lambda_\gamma$.

Anomalous TGCs affect both the total cross-section and the production angular distributions. Moreover, the relative contributions of each helicity sate would be modified, which in turn affect the angular distributions of the $W$ decay products.

\begin{figure}[h]
  \centering
  \includegraphics[width=8cm,height=8cm,angle=180]{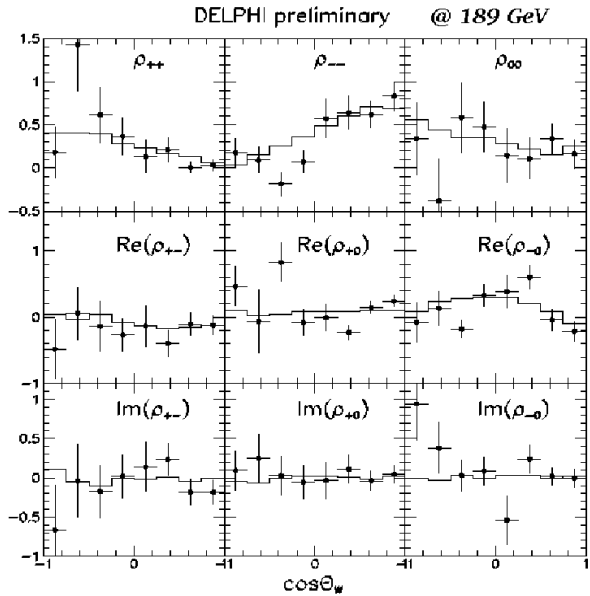}
  \caption{The SDM elements for the $W^+W^-$ production.}\label{fig:sdm}
\end{figure}

All the four collaborations use the event rate information in the three decay channels (hadronic, semileptonic and fullyleptonic) to measure the values of the TGCs, together with the event shape.
Different methods are used to analyse $WW$ events in order to extract the TGCs~\footnote{The LEP experiments are also computing the CP-violating TGCs.}: multidimensional phase-space fit or optimal observables~\cite{optimal}. Then the distributions are compared with the expectations relative to different values of the parameters, as obtained from fully simulated $WW$ Monte Carlo events.
A new method consists in the Spin Density Matrix (SDM) of which the elements (see figure~\ref{fig:sdm}) are observables directly related to the polarisation of the $W$~\cite{gounaris}. The comparison of the SDM elements with the theoretical predictions allows a model independent test of the TGCs.

There is no LEP combination since last summer because the experiments are waiting for a new $WW$ generator including higher order corrections. With respect to the old generators, $\mathcal O(\alpha)$ Monte Carlos predict a lower $WW$ cross-section (2.5\%) and a sizeable change in the slope of the $\cos \theta_W$ distributions (2\%). 
Only ALEPH has preliminary results~\cite{aleph}, listed in table~\ref{table:tgc}, including these higher order corrections.

\begin{table}[h]
\begin{center}
\begin{tabular}{|c||c|c|c|}
  \hline
  coupling & $\Delta g_1^Z$ & $\Delta \kappa_\gamma$ & $\lambda_\gamma$ \\
  \hline
  fit result  & $0.015^{+0.035}_{-0.032}$ & $-0.020^{+0.078}_{-0.072}$ & $-0.001^{+0.034}_{-0.031}$ \\
  \hline
\end{tabular}
\caption{ALEPH charged current TGC results including $\mathcal O(\alpha)$ corrections.}\label{table:tgc}
\end{center}
\end{table}

\subsection{Other channels}

The Single $W$, that is $We\nu$ final state, also gives information on the $WW\gamma$ vertex. The signature is one electron lost in the beam pipe and missing transverse momentum coming from the neutrino. The $We\nu$ channel has the same sensitivity to $\Delta \kappa_\gamma$ than the $WW$ channel and is also sensitive to $\lambda$.

The Single $\gamma$, that is $\nu\bar\nu\gamma$, concerns the $WW\gamma$ vertex only. The signature is a high energy photon isolated in the detector. The sensitivity to the couplings is about ten times lower than the $WW$ sensitivity.

\section{Quartic gauge couplings}\label{sec:qgc}

Four quartic gauge boson vertices are predicted in the SM with fixed couplings, $W^+W^-W^+W^-$, $W^+W^-Z^0Z^0$, $W^+W^-Z^0\gamma$ and $W^+W^-\gamma\gamma$, but their contributions to processes studied at LEP2 are negligible.
On the other hand, anomalous contributions to effective QGCs arising from physics beyond the SM could lead to measurable effects.

The formalism of anomalous QGCs involving at least one photon leads to the ``genuine'' QGC operators of dimension 6 after neglecting operators leading to triple gauge couplings.
Three anomalous QGCs, $a_0$, $a_c$ and $a_n$ are considered. $a_0$ and $a_c$ are CP-conserving and contribute to the $WW\gamma\gamma$ and $ZZ\gamma\gamma$ vertices~\cite{belanger}, whereas $a_n$ is CP-violating and contributes to the $WWZ\gamma$ vertex only~\cite{eboli}.

\begin{figure}[h]
  \centering
  \includegraphics[width=3cm,height=10cm,angle=-90]{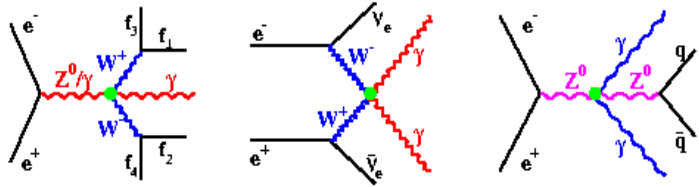}
  \caption{The three anomalous vertices $WWZ\gamma$, $WW\gamma\gamma$ and $ZZ\gamma\gamma$.}\label{fig:qgc}
\end{figure}

\begin{figure}
  \begin{minipage}[b]{.45\linewidth}
  \centering \includegraphics[width=5cm,height=5cm,angle=180]{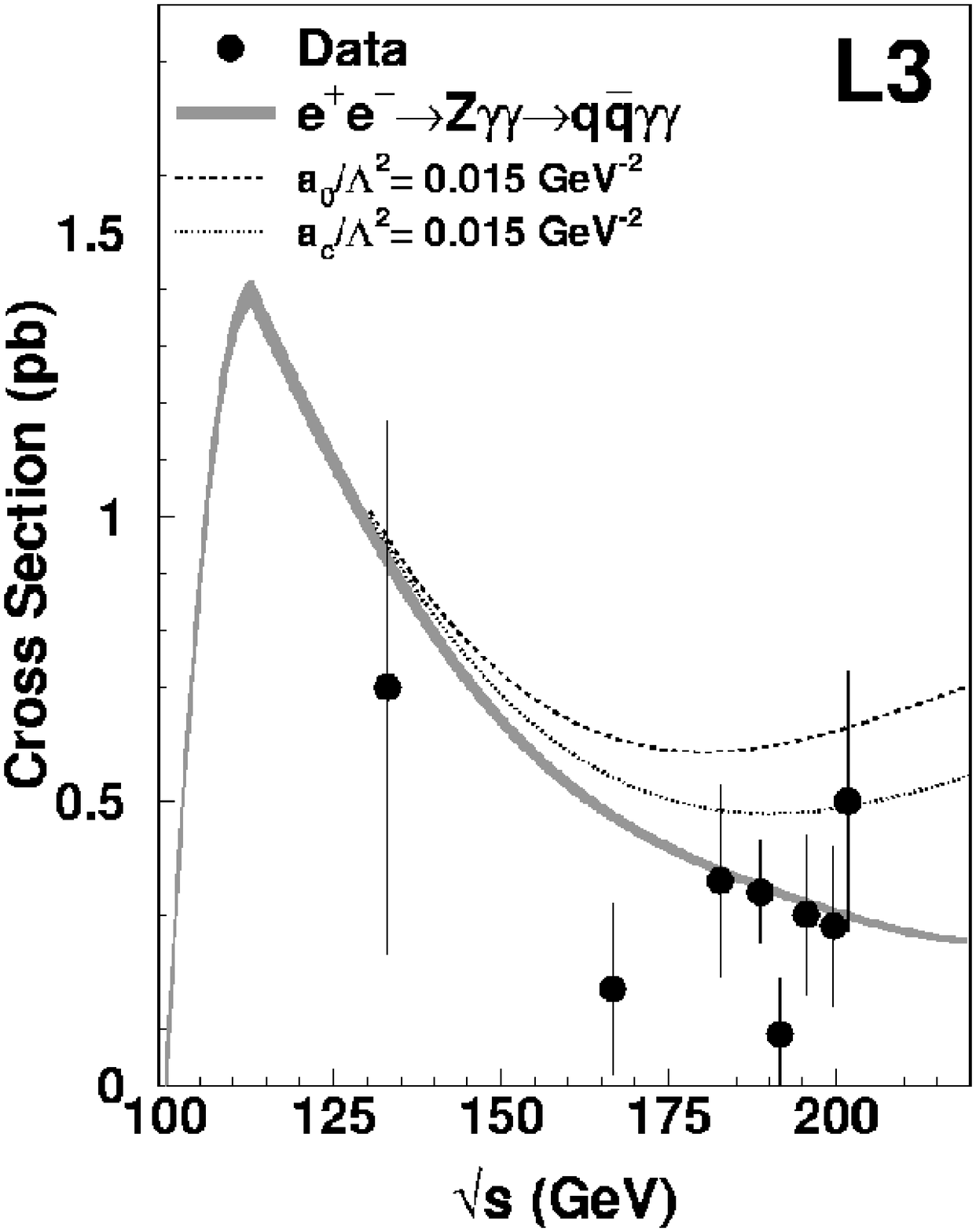}
  \caption{$Z\gamma\gamma$ cross-section with anomalous contributions from $a_0$ and $a_c$.}\label{fig:qgc-l3}
  \end{minipage}
  \begin{minipage}[b]{.45\linewidth}
  \centering \includegraphics[width=6cm,height=5cm,angle=180]{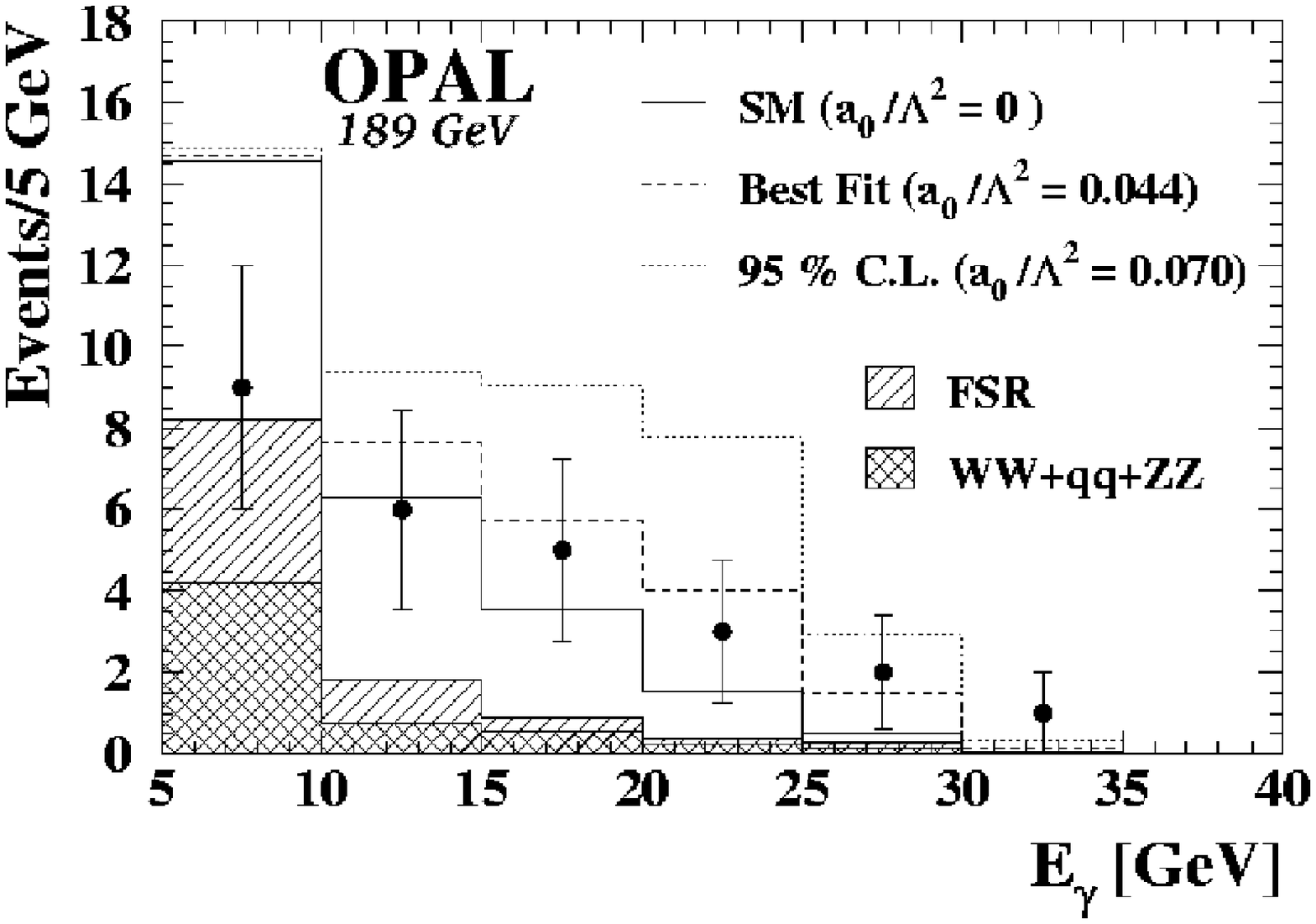}
  \caption{Energy distribution of the photon for the $WW\gamma$ channel.}\label{fig:qgc-opal}
  \end{minipage}
\end{figure}

Limits on $a_i / \Lambda^2 (i=0,c,n)$, with $\Lambda$ the energy scale where this new physics is supposed to appear, are obtained by comparing the cross-section (see figure~\ref{fig:qgc-l3}) and kinematic distributions (see figure~\ref{fig:qgc-opal}) of the $WW\gamma$, $\nu \bar\nu \gamma\gamma$ and $Z\gamma\gamma$ final states with the SM predictions.
The results are listed in table~\ref{table:qgc}~\cite{lepgc}~\footnote{There is no combination of neutral and charged QGCs because under more general theoretical hypothesis, effects in neutral QGCs may be different from charged QGCs ($a_i^Z \ne a_i^W$).}.

The $WW\gamma$ channel is analysed in the semileptonic and hadronic channels where the standard $WW$ selection is applied in addition to the search of a high energy and isolated photon.
A cut on low polar angles for the photon reduces the initial state radiations and a sharp mass window for the dijet reduces the final state radiations. The $W$ radiations are negligible.

The second charged anomalous QGC comes from the $\nu \bar\nu \gamma\gamma$ channel where two acoplanar photons are expected. A cut on missing mass is used to  reduce the $Z\gamma\gamma$ background.

The analysis concerning the neutral process leading to the $Z\gamma\gamma$ channel~\footnote{The $Z\gamma\gamma$ channel is not combined because the different generators used give different predictions.} is based on the search of high energy and isolated photons in hadronic events.

\begin{table}[h]
\begin{center}
\begin{tabular}{|c|c|c|}
  \hline
  parameter $[{\rm GeV}^{-2}]$ & vertex & 95\% CL \\
  \hline
  $a_0^W / \Lambda^2$ & $WW\gamma + \nu \bar\nu \gamma\gamma$ & [ -0.018 , 0.018 ] \\
  $a_c^W / \Lambda^2$ & $WW\gamma + \nu \bar\nu \gamma\gamma$ & [ -0.033 , 0.047 ] \\
  $a_n / \Lambda^2$ & $WW\gamma$ & [ -0.17 , 0.15 ] \\
  \hline
\end{tabular}
\caption{95\% CL combined LEP QGC results.}\label{table:qgc}
\end{center}
\end{table}

\section{Neutral TGCs}\label{sec:ngc}

The SM does not predict any direct coupling between the neutral gauge bosons themselves, but there are two possible anomalous vertices in the neutral sector~\cite{hagiwara}. Each vertex is parametrized by the most general Lorentz and $U(1)_{em}$ invariant lagrangian plus Bose symmetry and requiring only one off-shell boson~\footnote{DELPHI is also looking at off-shell couplings~\cite{off-shell}, canceling the condition of two outgoing on-shell bosons. This leads to 44 new couplings that are related to the $f^V_i$ and $h^V_i$ couplings.}.
These $ZZZ$, $ZZ\gamma$ and $Z\gamma \gamma$ vertices are all forbidden at tree level in the SM and have unobservably small values through loops.

\begin{figure}[h]
  \begin{tabular}{p{6cm}}
$Z\gamma$ final states are sensitive to possible contributions from anomalous $Z\gamma \gamma$ and $Z\gamma Z$ vertices which are parametrized by height couplings: $h^V_1$,  $h^V_2$ (CP-violating) and $h^V_3$, $h^V_4$ (CP-conserving) with $V = Z^* , \gamma^*$.
$Z\gamma$ events are investigated in $q\bar{q}\gamma$ and $\nu \bar{\nu}\gamma$ decay products where a hard visible photon and two jets or large missing energy and momentum for the neutrinos are searched for.
The $h^V_i$ couplings are fitted from the event rate (see figure~\ref{fig:ngc}), the angular and energy distributions of the photon. The results are shown in table~\ref{table:ngc}~\cite{lepgc} and figure~\ref{fig:res-ngc}.
  \end{tabular}
  \begin{minipage}{.45\linewidth}
  \centering   \includegraphics[width=5cm,height=5cm,angle=180]{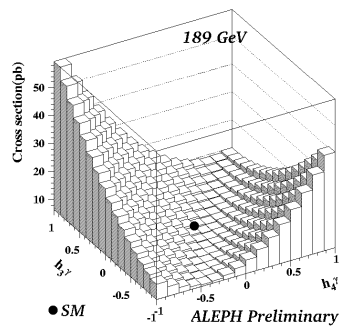}
  \caption{Variation of the $Z\gamma$ cross-section with respect to $(h_3^\gamma , h_4^\gamma)$.}\label{fig:ngc}
  \end{minipage}
\end{figure}

\begin{figure}[h]
\begin{center}
\begin{tabular}{ccc}
  \includegraphics[width=3.2cm,height=3.2cm]{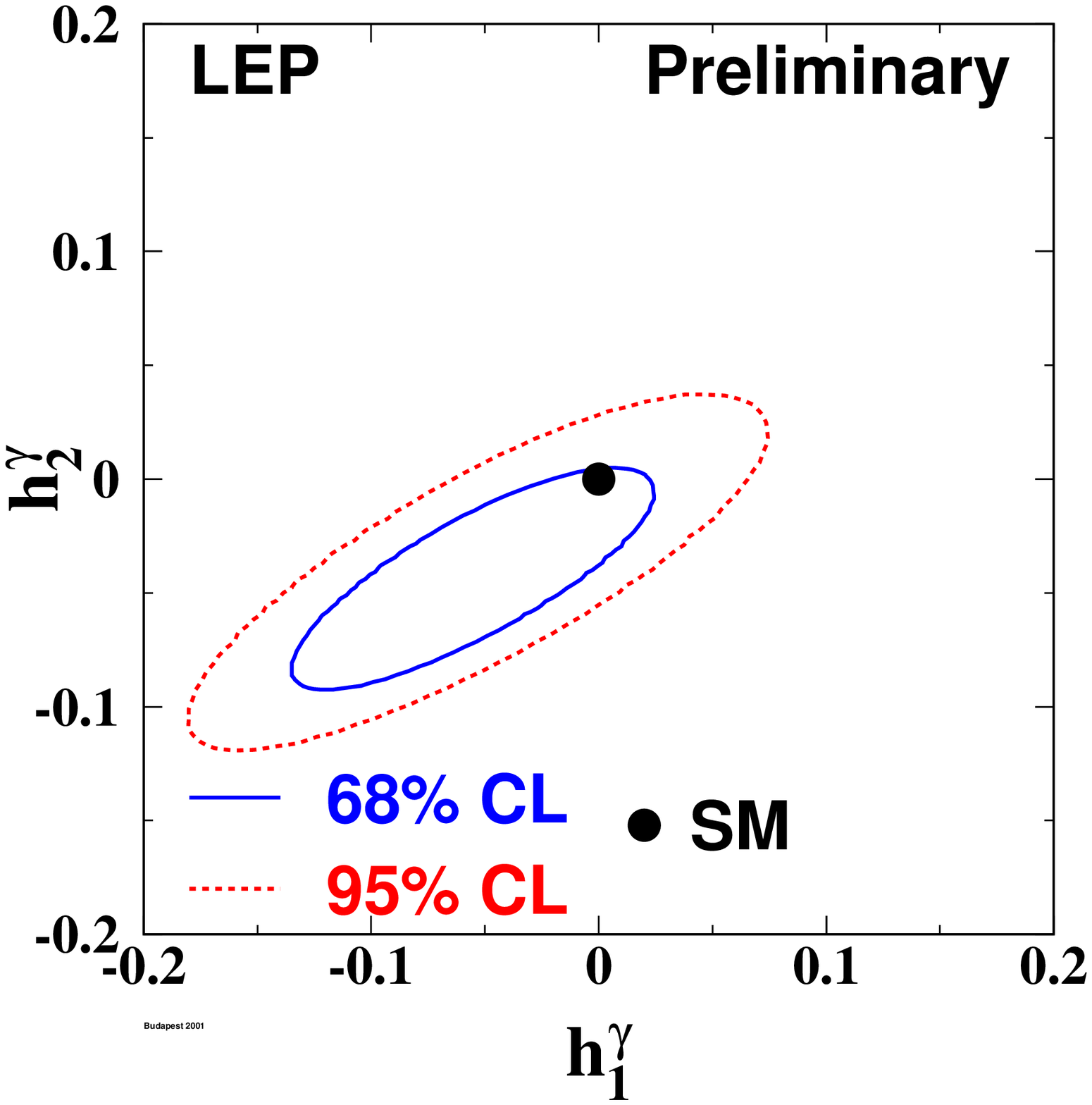} & \includegraphics[width=3.2cm,height=3.2cm]{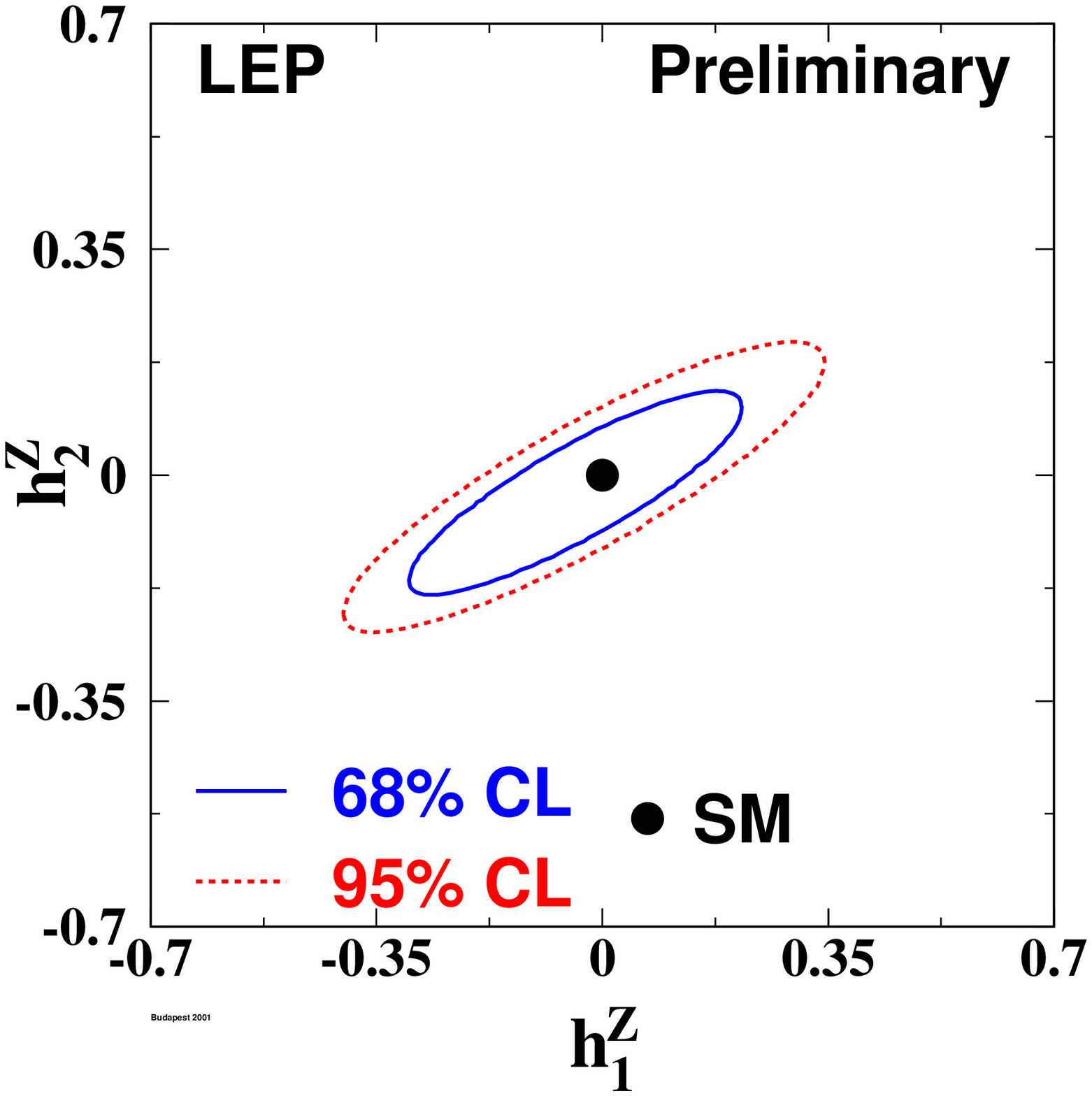} & \includegraphics[width=3.2cm,height=3.2cm]{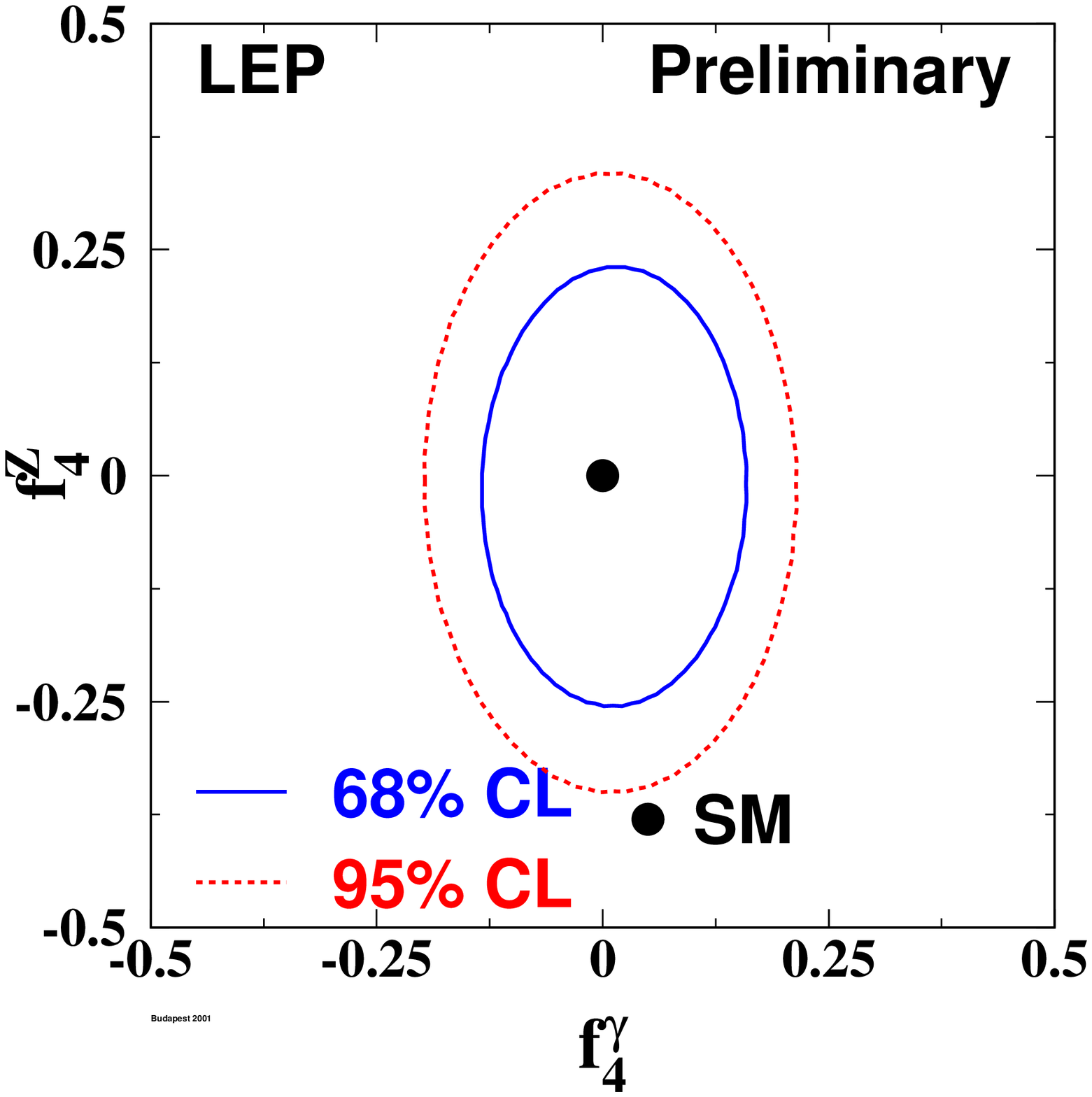} \\
\includegraphics[width=3cm,height=3cm]{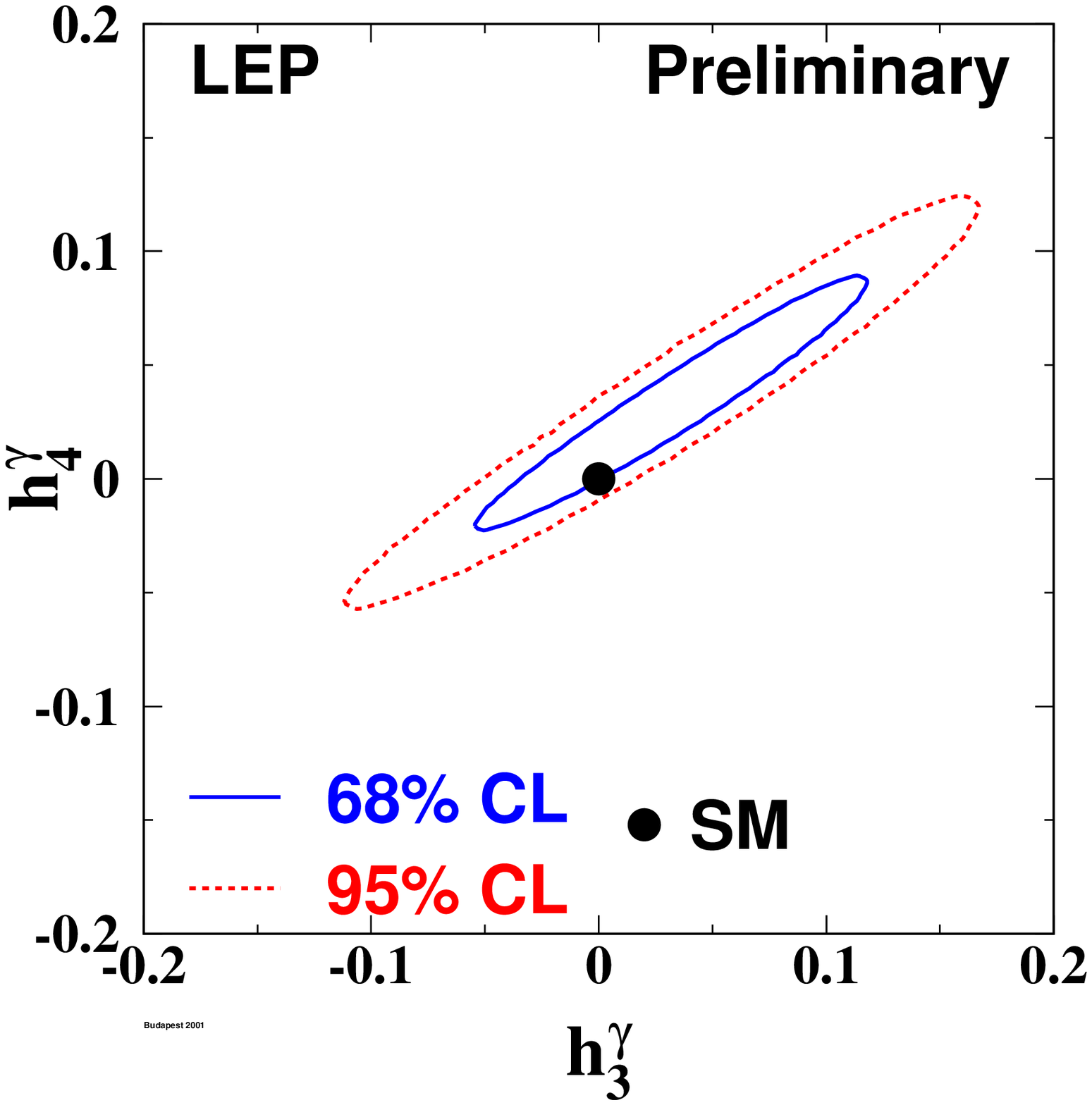} & \includegraphics[width=3cm,height=3cm]{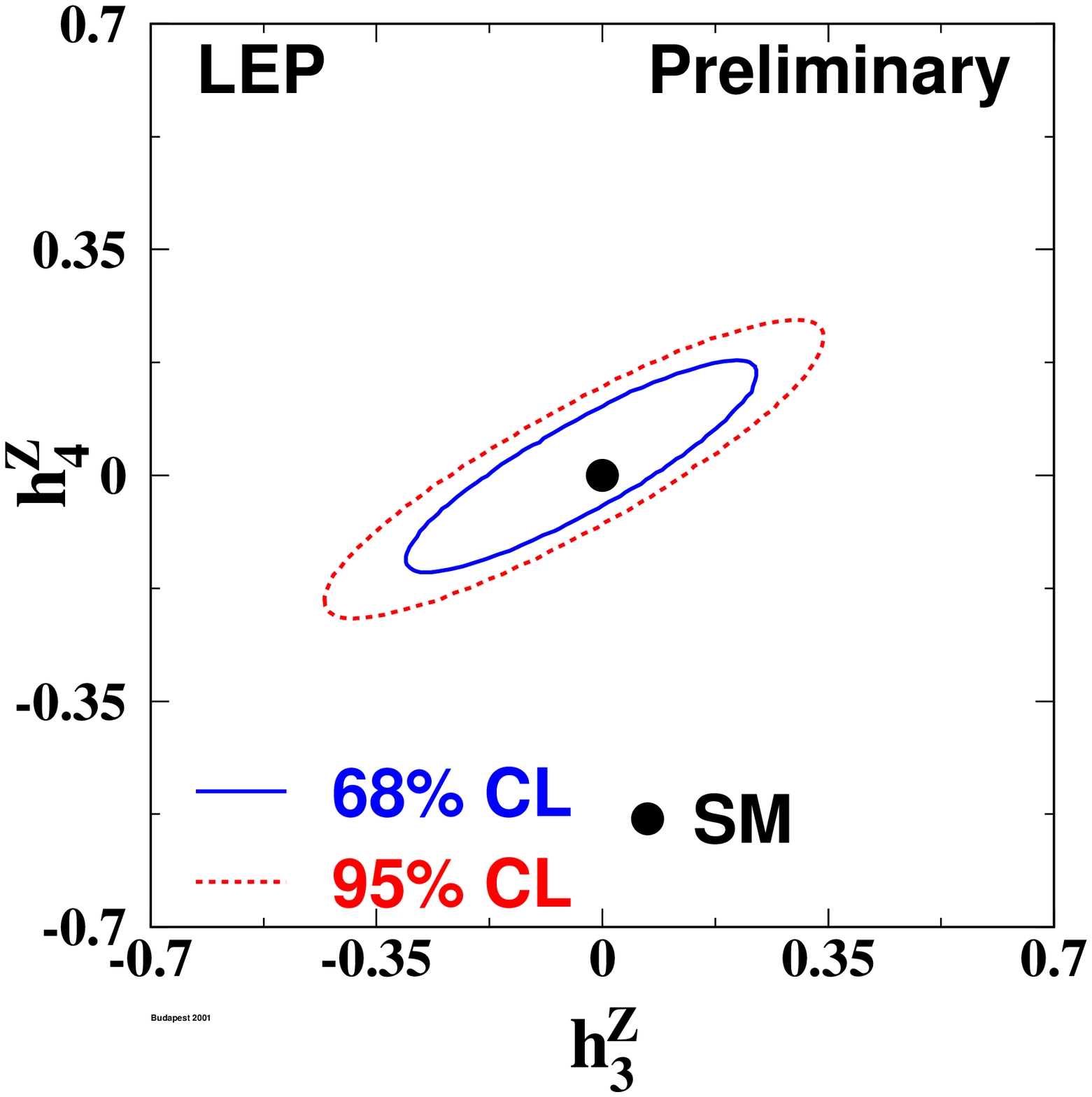} & \includegraphics[width=3cm,height=3cm]{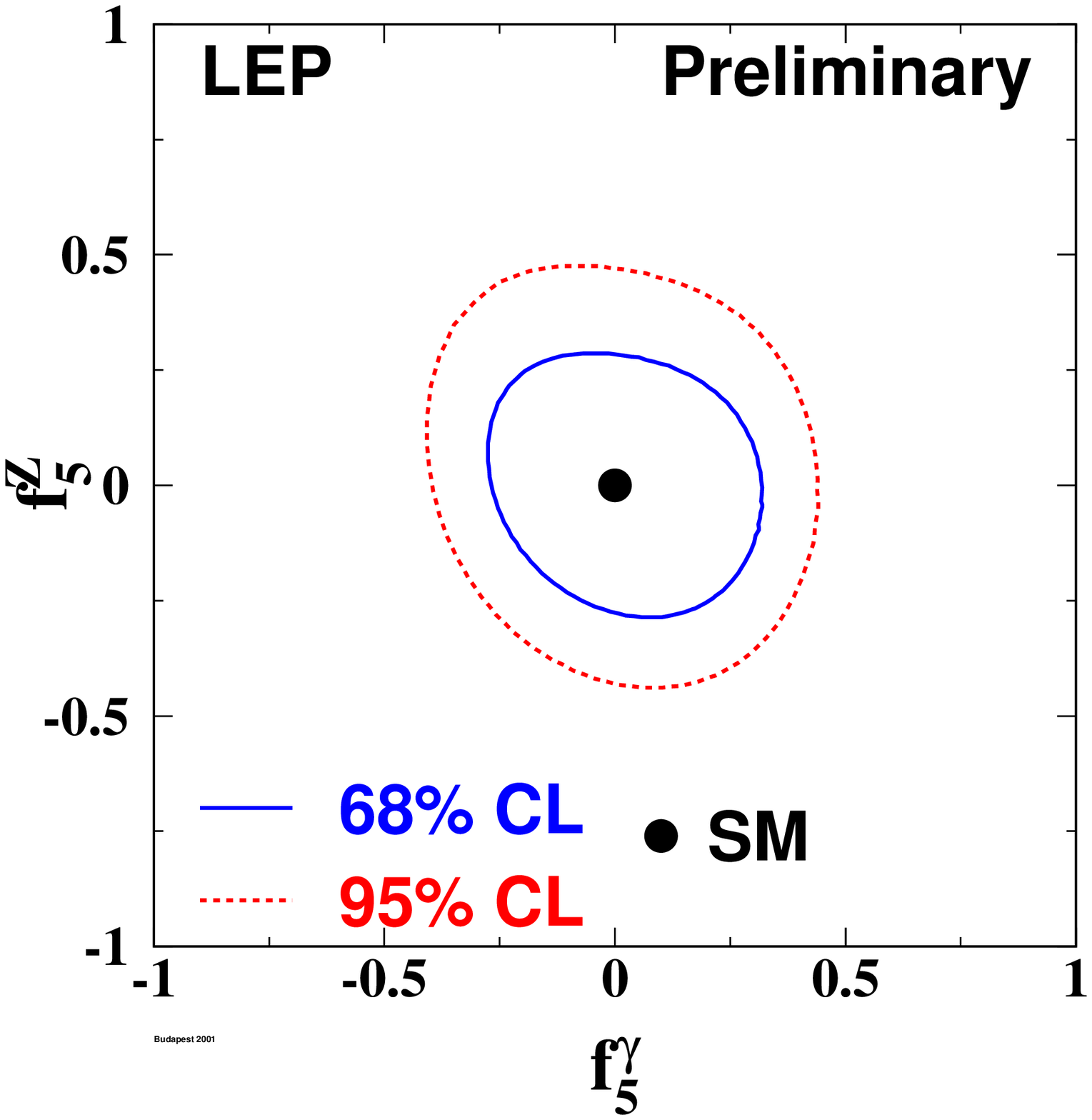} \\
\end{tabular}
\end{center}
  \caption{2D fit contours for combined LEP neutral current TGC.}\label{fig:res-ngc}
\end{figure}

Similarly to $Z\gamma$, $Z$-pairs can be used to constrain couplings related to the anomalous $ZZ\gamma$ and $ZZZ$ vertices. There are four couplings: $f^V_4$ (CP-violating) and $f^V_5$ (CP-conserving) with $V = Z^* , \gamma^*$.
The five visible decay channels are investigated: $Z \to q\bar{q}q\bar{q}$, $Z \to q\bar{q}\nu \bar{\nu}$, $Z \to q\bar{q}l^+l^-$, $Z \to l^+l^-\nu \bar{\nu}$ and  $Z \to l^+l^-l^+l^-$ with an expected branching ratio of 49\%, 28\%, 14\%, 4\% and 1\% respectively.
The Z-pair event rate, as well as angular distributions (mainly $\cos \theta_Z$) are used to constrain the values of the $f^V_i$ couplings. The results are summarized in table~\ref{table:ngc}~\cite{lepgc} and figure~\ref{fig:res-ngc}.

\begin{table}[h]
\begin{center}
\begin{tabular}{|cc|cc||cc|}
  \hline
  $h^{\gamma}_i$ & 95\% CL & $h^Z_i$ & 95\% CL & $f^V_i$ & 95\% CL \\
  \hline
  $h^{\gamma}_1$ & [-0.056,0.055] & $h^Z_1$ & [-0.128,0.126] & $f^{\gamma}_4$ & [-0.17,0.19] \\
  $h^{\gamma}_2$ & [-0.045,0.025] & $h^Z_2$ & [-0.078,0.071] & $f^Z_4$ & [-0.31,0.28] \\
  $h^{\gamma}_3$ & [-0.049,0.008] & $h^Z_3$ & [-0.197,0.074] & $f^{\gamma}_5$ & [-0.36,0.40] \\
  $h^{\gamma}_4$ & [-0.002,0.034] & $h^Z_4$ & [-0.049,0.124] & $f^Z_5$ & [-0.36,0.39] \\
  \hline
\end{tabular}
\caption{95\% CL combined LEP neutral current TGC results.}\label{table:ngc}
\end{center}
\end{table}

\section{Conclusions}\label{sec:conclusions}

Values and limits for anomalous triple and quartic gauge couplings in $e^+e^-$ collisions at centre of mass energies up to 208 GeV have been presented.
The Standard Model predictions are in well agreement with data.
The precision in the measurement of the gauge boson couplings has exceeded the expectations prior to the LEP2 startup.
The full potential of the LEP2 data is not yet fully exploited everywhere and future improvements of the combined LEP results can be expected.

\section*{Acknowledgments}

I would like to thank my ALEPH colleagues for interesting discussions about the physics of $W$ and $Z$ bosons.
Moreover, the warm hospitality of the 10th Lomonosov Conference has been well appreciated.

\end{document}